\documentclass[letterpaper, 10 pt, conference]{ieeeconf}  

\IEEEoverridecommandlockouts                              
\usepackage{hyperref}
\hypersetup{
    colorlinks=false,       
    linkcolor=red,          
    citecolor=green,        
    filecolor=magenta,      
    urlcolor=cyan           
}

\usepackage{amsmath,amssymb}
\usepackage{graphicx}
\usepackage{dcolumn}
\usepackage{amscd}
\usepackage{ifthen}
\usepackage{tikz}
\usepackage{pgfplots}
\usepackage{subfigure}

\usetikzlibrary{pgfplots.groupplots}

\newcommand\PR[1]{\ensuremath{{\mathrm{P}}\!\left[#1\right]}}
\newcommand\comp[1]{ \bar #1}

\newcommand\Tex{}
\newcommand\EX[2][\Tex]{
\ifthenelse{\equal{#1}{}}{{\mathbb E}\!\left[#2\right]}{\ensuremath{{\mathbb E}_{#1}\left[ #2\right]}}}

\newcommand\Var[2][\Tex]{
\ifthenelse{\equal{#1}{}}{{\mathrm{Var} }[#2]}{\ensuremath{\mathrm{Var}_{#1}\left[ #2\right]}}}

\newcommand\ignore[1]{}

\newcommand\defeq{:=}

 \newcommand{\mbf}[0]{\mathbf }

\newcommand\norm[2][\Tnorm]{\ensuremath{{\left\Vert #2 \right\Vert}_{#1}}}

\newcommand\Tinnerprod{}
\newcommand{\innerprod}[3][\Tinnerprod]{\ifthenelse{\equal{#1}{}}{\ensuremath{\left<#2,#3\right>}}{\ensuremath{\left<#2,#3\right>_{#1}}}}

\newcommand\vect[1]{\mathbf #1}

\newcommand{\reals}{\mathbb R} 
 
\newtheorem{definition}{Definition}

\newtheorem{theorem}{Theorem}

\newtheorem{proposition}{Proposition}

\newcommand{\pinv}[1]{  {#1}^{ \dagger } } 

\newcommand{\herm}[1]{{#1}^H} 

\renewcommand{\d}{d} 

\newcommand{\A}{\vect{A}}
\renewcommand{\a}{\vect{a}}

\newcommand{\e}{\vect{e}}

\renewcommand{\v}{\vect{v}}

\newcommand{\x}{\vect{x}}

\newcommand{\y}{\vect{y}}
\renewcommand{\r}{\vect{r}}

\newcommand{\I}{\vect{I}}

\renewcommand{\P}{\vect{P}}

\renewcommand{\S}{\mathcal S}

\newcommand{\sparsity}{s}
\newcommand{\spark}{\text{spark}}

\renewcommand{\d}{d}

\newcommand{\POPT}{\text{POPT}}
\newcommand{\LOPT}{\text{LOPT}}

\title{\LARGE \bf
Joint Sparsity with Different Measurement Matrices
}

\author{ \parbox{3 in}{\centering Reinhard Heckel and Helmut B\"olcskei\\
         Dept. of IT \& EE, ETH Zurich, Switzerland \\
         {\tt\small \{heckel,boelcskei\}@nari.ee.ethz.ch}}
}

\begin{document}

\maketitle
\thispagestyle{empty}
\pagestyle{empty}

\begin{abstract}
We consider a generalization of the multiple measurement vector (MMV) problem, where the measurement matrices are allowed to differ across measurements. This problem arises naturally when multiple measurements are taken over time, e.g., and the measurement modality (matrix) is time-varying. 
 We derive probabilistic recovery guarantees showing that---under certain (mild) conditions on the measurement matrices---$\ell_2/\ell_1$-norm minimization and a variant of orthogonal matching pursuit fail with a probability that decays exponentially in the number of measurements. This allows us to conclude that, perhaps surprisingly, recovery performance does not suffer from the individual measurements being taken through different measurement matrices. What is more, recovery performance typically benefits (significantly) from diversity in the measurement matrices; we specify conditions under which such improvements are obtained. These results continue to hold when the measurements are subject to (bounded) noise. 
\end{abstract}

\section{Introduction}

An interesting generalization of the sparse signal recovery problem as studied, e.g., in 
\cite{donoho_compressed_2006,donoho_optimally_2003,elad_sparse_2010},  
is the so-called multiple measurement vector (MMV) problem \cite{cotter_sparse_2005,chen_theoretical_2006,tropp_algorithms2_2006,tropp_algorithms_2006}. Application areas of the MMV problem include neuromagnetic imaging, array processing, and nonparametric spectral analysis of time series \cite{cotter_sparse_2005}. The MMV problem is formalized as follows: Given the vectors $\x^{(0)}, ..., \x^{(\d-1)}$, that share the sparsity pattern $\S$, i.e., the entries of $\x^{(0)}, ..., \x^{(\d-1)}$ are equal to zero on $\comp{\S}$, we want to recover the $\x^{(i)}$ from the noisy measurements $\y^{(i)} = \A \x^{(i)} + \e^{(i)}, i=0,...,\d-1$, where the $\e^{(i)}$ are noise vectors and the measurement matrix $\A \in \reals^{m\times n}$ is assumed known. For the noiseless case, i.e., $\e^{(i)}=\vect{0}$, for all $i$, 
it was shown in \cite{chen_theoretical_2006, davies_rank_2012} that the program
\newcommand{\PZ}{\text{(P0-MMV)}}
\[
\PZ  \,
\begin{cases} 
		\text{minimize} 	& |\S| \\
		\text{subject to} 	& \y^{(i)} \!\!=\! \A \x^{(i)},\quad  i=0,..,\d-1
\end{cases}
\]
recovers all $\x^{(0)},...,\x^{(\d-1)}$ with $\text{rank}{[\x^{(0)}...\; \x^{(\d-1)}]} = K$
if and only if
\begin{equation}
|\S| <   \frac{\spark(\A)-1+K}{2}
\label{eq:chenhuouniqeness}
\end{equation}
where $\spark(\A)$ is the cardinality of the smallest set of linearly dependent columns of $\A$ \cite{donoho_optimally_2003}. The  threshold \eqref{eq:chenhuouniqeness} constitutes a potentially significant improvement over the well-known $\spark(\A)/2$-threshold \cite{donoho_optimally_2003} for the single measurement vector (SMV) case, i.e., for $\d=1$. 
Necessity of the threshold \eqref{eq:chenhuouniqeness} shows that when asking for recovery of \emph{all} sets of vectors $\x^{(0)},...,\x^{(\d-1)}$, including linearly dependent collections, multiple measurements do not result in an improvement in the recovery threshold over the SMV case. 
It is therefore sensible to ask whether performance improvements can be expected for ``typical'' $\x^{(0)},...,\x^{(\d-1)}$. 
Since $\PZ$ is NP-hard \cite{davis_adaptive_1997}, this question is usually posed with the proviso that computationally efficient algorithms such as $\ell_2/\ell_1$-norm minimization or a variant of orthogonal matching pursuit (OMP)  \cite{chen_theoretical_2006,tropp_algorithms2_2006,tropp_algorithms_2006}  should be used for recovery. 
Indeed, a corresponding probabilistic performance analysis carried out in \cite{eldar_average_2010,gribonval_atoms_2008} 
shows that multiple measurements yield significant improvements in recovery performance over the SMV case. 

In practical applications the measurement matrix (modality) often changes across measurements, e.g., when measurements are taken over time and the underlying measurement modality exhibits characteristics that vary over time. 
It is therefore natural to ask whether improvements thanks to multiple measurements depend critically on the measurements all being taken through the same measurement matrix $\A$. 
We answer this question by considering the following modification of the MMV problem, termed generalized MMV (GMMV) problem henceforth: 
Given the vectors $\x^{(0)}, ..., \x^{(\d-1)}$, that share the sparsity pattern $\S$, recover the $\x^{(i)}$ from the (possibly noisy) measurements 
\begin{equation}
\y^{(i)} = \A^{(i)} \x^{(i)}   + \e^{(i)}, \quad i=0,...,\d-1
\label{eq:matrixvaries_1}
\end{equation}
assuming knowledge of the measurement matrices $\A^{(i)} \in \reals^{m\times n}$. Here, the $\e^{(i)}$ are noise vectors.

The GMMV problem also occurs in the recovery of sparse signals that lie in the union of shift-invariant subspaces  \cite{lu_theory_2008,eldar_compressed_2009}, as detailed in an extended version of this paper \cite{heckel_generalized_2012}.

As the MMV problem is a special case of the GMMV problem, obtained by setting $\A^{(i)} = \A$, for all $i$, it follows immediately that, for general $\A^{(i)}$, a worst-case (with respect to the $\x^{(i)}$) analysis reveals no improvements resulting from multiple measurements.

\paragraph*{Contributions}

The main theme of this paper is a probabilistic (with respect to the $\x^{(i)}$) performance analysis of an $\ell_2/\ell_1$-norm based recovery algorithm, called LOPT, and a variant of OMP, called MOMP, for deterministic measurement matrices $\A^{(i)}$. For the noiseless case, under very general conditions on the $\A^{(i)}$, we find that the failure probability of LOPT and MOMP decays exponentially in the number of measurements $\d$. We show that, perhaps surprisingly, having different measurement matrices $\A^{(i)}$ can lead to (substantial) performance improvements over the MMV case $\A^{(0)} = ... = \A^{(\d-1)}$. What is more, these improvements are obtained under very mild ``isometry'' conditions on the $\A^{(i)}$. Furthermore, we show that our results continue to hold when the measurements are subject to bounded noise. 

The probabilistic model on the $\x^{(i)}$~we use is more general than that employed in \cite{eldar_average_2010, gribonval_atoms_2008} for the MMV case. Particularizing our results to the MMV case 
therefore yields generalizations of the main results in \cite{eldar_average_2010, gribonval_atoms_2008}. For the noisy case our result for LOPT is new, even in the MMV case.

We note that the GMMV problem can be cast as a block-sparse problem \cite{eldar_block-sparse_2010}, which in turn is contained in the model-based \cite{baraniuk_model-based_2010} setting. However, formulating the GMMV problem as a block-sparse (or model-based) problem, and applying the corresponding recovery results available in the literature yields worst-case recovery conditions only. 

In terms of mathematical tools, we note that the proofs of our main results,  provided in \cite{heckel_generalized_2012}, consist of two steps. First, we derive conditions for LOPT and MOMP to succeed and then we use concentration of measure results to show that these conditions are satisfied with high probability, provided that mild conditions on the $\A^{(i)}$ are satisfied. While the proofs in \cite{eldar_average_2010, gribonval_atoms_2008} follow these two general steps as well, the technical specifics are quite different. 
Concretely, the more general probabilistic model for the $\x^{(i)}$ requires the use of concentration of measure results that are more general than those employed in \cite{eldar_average_2010, gribonval_atoms_2008}. In addition,  
our recovery conditions are new, and, in particular in the noisy case, non-trivial to derive. 

\paragraph*{Notation} We use lowercase boldface letters to denote column vectors, e.g., $\mbf x$, and uppercase boldface letters to designate matrices, e.g., $\mbf A$. For a vector $\x$, $[\x]_q$ and $x_q$ denote the $q$th entry. 
For the matrix $\A$,  $\pinv{\A}$ is its pseudo-inverse and $\norm[2\to 2]{\A} \defeq \max_{\norm[2]{\v} = 1  } \norm[2]{\A \v}$ its spectral norm. 
The superscript $\herm{}$ stands for  Hermitian transposition.   
For the set $\S$, $|\S|$ is its cardinality and $\comp{\S}$ stands for its complement in $\{0,...,n-1\}$. 
We say that a random variable $x$ is standard Gaussian, if it is of zero-mean and unit variance; $x$ is standard complex Gaussian if $x = x_R + j x_I$, where $ x_R,x_I$ are i.i.d.~Gaussian with mean zero and variance $1/2$. 

\section{Problem formulation}
The formal statement of the problem we consider is as follows. 
Suppose we observe the $m$-dimensional vectors 
\begin{equation}
\y^{(i)} = \A^{(i)} \x^{(i)} + \e^{(i)}, \quad i=0,...,\d-1
\label{eq:noisymeasurement}
\end{equation}
where the $\e^{(i)} \in \reals^{m}$ account for (unknown) noise, the $\x^{(i)} \in \reals^n$, $n>m$, share the sparsity pattern $\S \subseteq \{0,...,n-1\}$, i.e., for each $\x^{(i)}$ the entries with index in $\comp{\S}$ are equal to zero, and the measurement matrices $\A^{(0)},...,\A^{(\d-1)} \in \reals^{m\times n}$ are known. 
We want to recover $\x^{(0)},...,\x^{(\d-1)}$ from the $\y^{(0)},...,\y^{(\d-1)}$. 

We first consider the noiseless case, i.e., $\e^{(i)} = \vect{0}$, for all $i$. 
Recovery can be accomplished by solving 
\newcommand{\PZGM}{\text{P0-GMMV}}
\[
(\PZGM)  \,
\begin{cases} 
		\text{minimize} 	& |\S| \\
		\text{subject to} 	& \y^{(i)} = \A^{(i)} \x^{(i)},\;  i=0,...,\d-1
\end{cases}
\]
which is, however, NP-hard \cite{davis_adaptive_1997}. 
Computationally efficient alternative recovery algorithms, with, however, weaker recovery guarantees,  are specified next. 
A convex relaxation of $\PZGM$ is given by
\[
(\LOPT) \;
\begin{cases}
\text{minimize }   \sum_{l=0}^{n-1}  \!\left( \sum_{i=0}^{\d-1}  \left| x_l^{(i)} \right|^2   \right)^{1/2}     \\
\text{subject to } \y^{(i)} = \A^{(i)} \x^{(i)}, \; i=0,...,\d-1.
\end{cases}
\]
Another alternative, which is an adaptation of OMP, and will be called MOMP, is defined as follows. 
MOMP iteratively builds up the joint support set of $\x^{(0)},...,\x^{(\d-1)}$. 
The algorithm is initialized by choosing the residuals in iteration $0$ as $\r^{(i)}_0 = \y^{(i)}, i=0,...,\d-1$, and the set of selected indices as $\S_0=\emptyset$. In the $p$th iteration ($p\geq 1$) we find the index 
\[
l_p = \arg \max_l   \sum_{i=0}^{\d-1} \left| \herm{(\a_l^{(i)})}  \r_{p-1}^{(i)}  \right|^2                      
\]
and update the set of selected indices by setting $\S_p = \S_{p-1} \cup \{ l_p \}$. The residuals are updated according to
\[
\r_p^{(i)} = \y^{(i)} - \A_{\S_p}^{(i)} \x_{\S_p}^{(i)} = (\I - \P_{\S_p}^{(i)}) \y^{(i)}, \quad i = 0,..., \d-1
\]
where $\A_{\S_p}^{(i)}$ is the matrix obtained from $\A^{(i)}$ by selecting the columns with indices in $\S_p$ and $\P_{\S_p}^{(i)} \defeq \A_{\S_p}^{(i)} \pinv{  ( \A_{\S_p}^{(i)} )}$ is the orthogonal projector onto the span of the columns in $\A_{\S_p}^{(i)}$. 
Both LOPT and MOMP are trivial generalizations of corresponding algorithms for the MMV case 
  \cite{cotter_sparse_2005,chen_theoretical_2006,tropp_algorithms2_2006,tropp_algorithms_2006}.

Proceeding to the noisy case, we assume that noise is bounded in the sense of
\begin{align}
\sum_{i=0}^{\d-1} \norm[2]{\e^{(i)}}^2  \leq \epsilon^2.
\label{eq:jointnoisconst}
\end{align}
As exact recovery of the $\x^{(i)}$ will, in general, no longer be possible, we will be content with ensuring that the estimates of the $\x^{(i)}$ are ``close'' to the true $\x^{(i)}$.  
The recovery algorithms we analyze in the noisy case are MOMP and a convex program closely related to LOPT, namely  
\begin{align*}
(\POPT) \;\, \text{minimize }    &\frac{1}{2} \sum_{i=0}^{\d-1} \norm[2]{\y^{(i)} -\A^{(i)} \x^{(i)}  }^2   \\
&+     \gamma  \sum_{l=0}^{n-1}  \!\left( \sum_{i=0}^{\d-1}  \left| x_l^{(i)} \right|^2   \right)^{1/2}
\end{align*}
which, for $\d=1$, is known as \emph{the lasso} \cite{tibshirani_regression_1996} in the statistics literature, and for $\d>1$, is a particular variant of \emph{the group lasso} \cite{yuan_model_2006}. The first term in the cost function of $\POPT$ accounts for the recovery error and the second term enforces sparsity; the parameter $\gamma>0$ controls the tradeoff between these two terms.

\section{Review of worst-case recovery results \label{sec:blockdefs}}
We briefly discuss worst-case recovery results for the GMMV problem.  Formulating the GMMV problem as a block-sparse recovery problem and evaluating the corresponding recovery conditions in \cite{eldar_block-sparse_2010} yields the following proposition.  
\begin{proposition}
Let $\S$ be the sparsity pattern of $\x^{(0)},...,\x^{(\d-1)}$ and assume that 
\begin{equation}
\max_{l\notin \S} \sum_{q}  \max_{i=0,...,d-1}  \left| [\pinv{(\A^{(i)}_\S)} \a_l^{(i)}]_q \right| < 1.
\label{eq:block_recovery_wccond}
\end{equation}
Then, $\LOPT$ and MOMP recover $\x^{(0)},...,\x^{(\d-1)}$ exactly from $\y^{(i)} = \A^{(i)} \x^{(i)},  i=0,...,\d-1$.
\label{thm:exactreccond}
\end{proposition}

For the MMV case, Proposition \ref{thm:exactreccond} reduces to \cite[Th. 3.1]{chen_theoretical_2006}. 
 Condition \eqref{eq:block_recovery_wccond} can be viewed as the GMMV-equivalent of the SMV-exact recovery condition, a standard recovery condition for $\ell_1$-minimization and OMP \cite{tropp_greed_2004}. 
 
An alternative recovery condition can be obtained by viewing the GMMV problem as separate SMV problems and requiring exact recovery for each of the resulting SMV problems. Following this route, based on the SMV exact recovery condition \cite[Th. A]{tropp_greed_2004}, we get that $\ell_1$-minimization and OMP applied individually to $\y^{(i)} = \A^{(i)} \x^{(i)}$ recover  $\x^{(0)},...,\x^{(\d-1)}$ correctly if 
\begin{equation}
\max_{l\notin \S} \max_{i=0,...,d-1}   \norm[1]{\pinv{(\A^{(i)}_\S)} \a_l^{(i)}} < 1.
\label{eq:cond_recallindividually}
\end{equation}
This is a slightly weaker condition than \eqref{eq:block_recovery_wccond}. 
Hence Proposition \ref{thm:exactreccond} does not predict any improvement of using LOPT or MOMP over treating the recovery problem as individual SMV problems (solved through $\ell_1$-minimization and/or OMP). 
\section{Main results}



We discuss the noiseless and the noisy case separately. 

\subsection{Recovery in the noiseless case \label{sec:exactrec}}

For the noiseless case the probabilistic model on the $\x^{(i)}$ is as follows:  For a given support set $\S \subseteq \{0,...,n-1\}$, we take the entries of the vectors $\x^{(0)}_\S\!,...,\x^{(\d-1)}_\S$ to be independent sub-Gaussian \cite{vershynin_introduction_2012}. 
\newcommand{\Kg}{K}
\begin{definition}
A zero-mean random variable is $\rho$-sub-Gaussian\footnote{Sub-Gaussian random variables are often equivalently defined through tail bounds or through moment bounds, see e.g. \cite{vershynin_introduction_2012}. The definition we chose is the most convenient for our purposes. 
}, with $\rho>0$, if its moment generating function satisfies 
\begin{equation}
\EX{e^{tx}} \leq e^{\rho t^2}.
\label{eq:subGmomgenfunc}
\end{equation}
\end{definition}

Sub-Gaussian random variables contain Gaussian and all bounded\footnote{The random variable $x$ is bounded if there exists an $M\geq0$ such that $\PR{|x|\leq M} =1$.}
random variables as special cases. 
We start with our main result for LOPT in the noiseless case. 
\begin{theorem}
Fix $\S \subseteq \{0,...,n-1\}$ with cardinality $\sparsity \defeq |\S|$,   
and take the entries of $\x^{(0)}_\S\!,...,\x^{(\d-1)}_\S  \in \reals^{\sparsity}$ to be i.i.d.~zero-mean $\rho$-sub-Gaussian with unit variance\footnote{This is w.l.o.g. as the entries of the $\x^{(i)}$ can be scaled to account for non-unit variance.}. 
 Assume that the measurement matrices $\A^{(0)}, ...,\allowbreak \A^{(\d-1)} \in \reals^{m\times n}$ satisfy 
\begin{equation}
\left(\frac{1}{d} \sum_{i=0}^{\d-1} \norm[2]{\pinv{(\A_\S^{(i)})} \a_l^{(i)}}^2  \right)^{1/2} 
\leq \alpha < 1, \, \text{ for all } \,l \notin \S
\label{eq:cond_prob_mixenormres}
\end{equation}
and 
\begin{equation}
\max_i \norm[2]{\pinv{(\A_\S^{(i)})} \a_l^{(i)}}  
\leq \gamma, \, \text{ for all } \,l \notin \S
\label{eq:cond_prob_mixenormres2}
\end{equation} 
for some $\gamma>0$, where $\a_l^{(i)}$ denotes the $l$th column of $\A^{(i)}$. 
Then, for every $\xi>0$ satisfying $\max \{ 1-32 e \rho,   \alpha^2 \}  < \xi^2 \leq  \alpha^2(1+32 e \rho)$, with probability at least
\begin{equation}
1-  (n-s)   \exp \! \left(   -  \d   \frac{  (\xi^2-\alpha^2)^2}{2^{11} e^2  \rho^2  \gamma^2  \alpha^2 } \right)   -  
\sparsity \exp \! \left(  -  \d  \frac{( 1-\xi^2)^2}{ 2^{11} e^2  \rho^2 }   \right)
\label{eq:prboundinmnmSG}
\end{equation}
$\LOPT$ applied to $\y^{(i)} = \A^{(i)} \x^{(i)},  i=0,...,\d-1$, recovers the correct solution $\x^{(0)},...,\x^{(\d-1)}$. 
\label{thm:mixednorm_prob_subg}
\label{thm:mixednorm_prob}
\end{theorem}

The main implication of Theorem \ref{thm:mixednorm_prob} is that, provided \eqref{eq:cond_prob_mixenormres} (and \eqref{eq:cond_prob_mixenormres2}) is satisfied, the probability that $\LOPT$ fails decays exponentially in the number of measurements $\d$. This has been shown before for the MMV case under the assumption of i.i.d.~Gaussian $\x_\S^{(i)}$ \cite[Th. 4.4]{eldar_average_2010}.  

 The constants in the exponents of \eqref{eq:prboundinmnmSG} can be improved (significantly) for certain distributions. For example, 
 when the entries of the $\x_\S^{(i)}$ are i.i.d.~standard Gaussian (note that a standard Gaussian is sub-Gaussian with $\rho=1/2$), the recovery probability   is at least   \cite{heckel_generalized_2012} 
\begin{equation}
1-  (n-s)    \exp \! \left( -d \frac{(\xi - \alpha )^2}{2 \gamma^2} \right)  -  
\sparsity \exp \! \left( -d \frac{ (1-\xi^2)^2}{4} \right). 
\label{eq:prboundinmnm}
\end{equation}

\paragraph*{Improvements over worst-case results}
First note that $\gamma$  in \eqref{eq:cond_prob_mixenormres2} can be chosen arbitrarily, hence \eqref{eq:cond_prob_mixenormres2} is not restrictive.
To see that the recovery condition \eqref{eq:cond_prob_mixenormres} is weaker than the worst-case recovery condition \eqref{eq:cond_recallindividually} (recall that \eqref{eq:cond_recallindividually} implies \eqref{eq:block_recovery_wccond}), we simply note that 
\begin{align*}
\left(\frac{1}{d} \sum_{i=0}^{\d-1} \norm[2]{\pinv{(\A_\S^{(i)})} \a_l^{(i)}}^2 \right)^{1/2} 
&\leq 
\left(\frac{1}{d} \sum_{i=0}^{\d-1} \norm[1]{\pinv{(\A_\S^{(i)})} \a_l^{(i)}}^2  \right)^{1/2}  \\
&\leq \max_{i=0,...,d-1} \norm[1]{\pinv{(\A_\S^{(i)})} \a_l^{(i)}}.
\end{align*}

\paragraph*{Improvements due to different measurement matrices}
Evaluating 
\eqref{eq:cond_prob_mixenormres} for the MMV case yields 
\begin{align}
\norm[2]{\pinv{(\A_\S)} \a_l}  \leq \alpha < 1, \text{ for all } \,l \notin \S.
\label{eq:cond_prob_mixenormresformmv}
\end{align}
Note that \eqref{eq:cond_prob_mixenormresformmv} is the recovery condition stated in \cite[Th. 4.4]{eldar_average_2010} and applying to the case where the entries of the $\x_\S^{(i)}$ are i.i.d.~Gaussian. 
Comparing \eqref{eq:cond_prob_mixenormresformmv} to \eqref{eq:cond_prob_mixenormres}, we see that in the GMMV case the measurement matrices have to satisfy $\norm[2]{\pinv{(\A^{(i)}_\S)} \a_l^{(i)}}^2  \leq \alpha^2$ only on average (i.e., across $i$). 
This essentially says that having different measurement matrices allows for some of them to be ``bad'' as long as the collection $\{\A^{(0)},...,\A^{(\d-1)}\}$ is good enough on average. In contrast, in the MMV case, the single measurement matrix $\A$ has to be ``good'' in the sense of \eqref{eq:cond_prob_mixenormresformmv}.

This can be nicely illustrated by way of an example. Suppose we are given a measurement matrix $\A$ which does not satisfy  \eqref{eq:cond_prob_mixenormresformmv} for all $\S \subseteq \{0,...,n-1\}$ with $|\S| \leq k$, for a given $k$, but does so on average over those $\S$.  
Now, take the matrices $\A^{(0)},...,\A^{(\d-1)}$ to be obtained independently by permuting the columns of $\A$. Then, if $\d$ is sufficiently large, with high probability \eqref{eq:cond_prob_mixenormres} will be satisfied for all $\S$ with $|\S| \leq k$.

We next state our recovery results for MOMP and  start by defining the following quantities, which are used to formulate ``local'' (i.e., pertaining to the (given) set $\S$) isometry conditions. These quantities were also used in \cite{eldar_average_2010,gribonval_atoms_2008} in the performance 
analysis of MOMP for the MMV case. 

For a given set $\S \subseteq \{0,...,n-1\}$, let 
\[
\delta_i(\S) = \norm[2\to2]{\herm{(\A_\S^{(i)})} \! \A^{(i)}_\S - \I}.
\]
Observe that 
\begin{equation*}
(1- \delta_i(\S))\norm[2]{\x_\S}^2   \leq  \norm[2]{\A_\S^{(i)}  \x_\S }^2  \leq (1+\delta_i(\S)) \norm[2]{\x_\S}^2  
\end{equation*}
for all $\x_\S \in \reals^\sparsity$. Define 
\[
\mu_i(\S) = \max \left\{   \max_{l\notin \S}\norm[2]{\herm{(\A_\S^{(i)})} \a^{(i)}_l },   \max_{l \in \S}\norm[2]{   \herm{(\A_{\S\setminus l}^{(i)})}   \a^{(i)}_l  }    \right\}
\]
and let $\delta_{\max}(\S)  = \max_i \delta_i(\S)$ and $\mu_{\max}(\S)  = \max_i \mu_i(\S)$.

\begin{theorem}
Fix $\S \subseteq \{0,...,n-1\}$ with cardinality $\sparsity\defeq |\S|$, let the measurement matrices $\A^{(0)}, ..., \A^{(\d-1)}\in \reals^{m\times n}$ have unit norm columns with $\mu_{\max}(\S)<1$ and $\delta_{\max}(\S) <1$, 
and let the entries of $\x^{(0)}_\S\!,...,\x^{(\d-1)}_\S  \in \reals^{\sparsity}$ be i.i.d.~zero-mean $\rho$-sub-Gaussian with unit variance. If
\begin{equation}
\frac{\sum_{i=0}^{\d-1}  \left(\frac{\mu_i(\S)}{1-\delta_i(\S)} \right)^2}{\sum_{i=0}^{\d-1}  \left( 1-   \frac{\mu_i^2(\S)}{1-\delta_i(\S)}  \right)^2  } \leq   \frac{(1-\beta)}{(1+\beta)} 
\label{eq:ompavreccond_nc}
\end{equation}
for $\beta$ with $0<\beta \leq 32 e \rho$, then MOMP applied to $\y^{(i)} = \A^{(i)} \x^{(i)}, i=0,...,\d-1$,  recovers the correct solution $\x^{(0)},...,\x^{(\d-1)}$ with probability at least
\begin{equation}
1- 2^\sparsity   (n +1 - \sparsity)  \exp \! \left(  - d \beta^2  \frac{c(\S,\A)}{2^{11} e^2\rho^2 } \right) 
\label{eq:bocorrrecprob}
\end{equation}
where $c(\S, \A)$ is a constant that depends on the $\A^{(i)}$, but is independent of $\d$. 
\label{thm:ompavcase}
\end{theorem}

\paragraph*{Remark}The constant $c(\S, \A)$ can be lower-bounded in terms of the $\mu_{i}(\S)$ and $\delta_{i}(\S)$, see \cite{heckel_generalized_2012}. 

The main implication of Theorem \ref{thm:ompavcase} is that, provided \eqref{eq:ompavreccond_nc} is satisfied, the probability that MOMP fails decays exponentially in the number of measurements $\d$. This has been shown before in \cite{eldar_average_2010, gribonval_atoms_2008} for the MMV case, under the assumption of i.i.d.~Gaussian $\x_\S^{(i)}$. 
 The implications of Theorem \ref{thm:ompavcase}  concerning improvements over the worst-case results and over the MMV case are as discussed above, for LOPT. 
Furthermore, as in the case of LOPT, Theorem \ref{thm:ompavcase} can be strengthened for certain distributions. For example, when the entries of $\x^{(0)}_\S\!,...,\x^{(\d-1)}_\S$ are i.i.d.~standard Gaussian, Theorem \ref{thm:ompavcase} holds with Condition \eqref{eq:ompavreccond_nc} replaced by
\begin{equation}
\frac{\sum_{i=0}^{\d-1}  \left(\frac{\mu_i(\S)}{1-\delta_i(\S)} \right)^2}{\sum_{i=0}^{\d-1}  \left( 1-   \frac{\mu_i^2(\S)}{1-\delta_i(\S)}  \right)^2  } \leq   \frac{(1-\beta)^2\,\varsigma^2}{(1+\beta)^2} 
\label{eq:ompavreccond_ncg}
\end{equation}
for $\beta>0$, where $\varsigma>1$ is a constant that tends to $1$ as $\d$ grows, and \eqref{eq:bocorrrecprob} replaced by 
\begin{equation}
1- 2^\sparsity \! \left(   (n - \sparsity) \exp \! \left(  - d \beta^2 c(\S, \A)   \right)    +  \exp \! \left(  -d  \beta^2  \varsigma^2 c(\S, \A)   \right)
 \right).
\label{eq:bocorrrecprobg}
\end{equation}

 We finally note that condition \eqref{eq:ompavreccond_nc} is slightly stronger than condition \eqref{eq:cond_prob_mixenormres} pertaining to $\LOPT$ \cite{heckel_generalized_2012}.


\subsection{The noisy GMMV problem} 

We next present our results for the noisy GMMV problem and start with the probabilistic analysis of $\POPT$. For the following result, we assume that the entries of $\x^{(0)}_\S\!,...,\x^{(\d-1)}_\S$ are i.i.d.~Rademacher random variables, i.e., they take on the values $+1$ and $-1$ with equal probability. We chose this model for convenience and note that similar results can be obtained for the sub-Gaussian case. 
The corresponding analysis is, however, much more cumbersome and does not yield additional insights. 

\begin{theorem}
Fix $\S \subseteq \{0,...,n-1\}$, with cardinality $\sparsity \defeq |\S|$,   
and take the entries of $\x^{(0)}_\S\!,...,\x^{(\d-1)}_\S  \in \reals^{\sparsity}$ to be i.i.d.~Rademacher. Suppose the measurement matrices $\A^{(0)}, ..., \A^{(\d-1)} \in \reals^{m\times n}$ satisfy conditions \eqref{eq:cond_prob_mixenormres}  and \eqref{eq:cond_prob_mixenormres2}   for $\alpha < 1$ and some $\gamma>0$. Suppose the noise level $\epsilon$ in \eqref{eq:jointnoisconst} and $\gamma$ satisfy
\begin{align}
\left( c_3 \epsilon + \gamma c_4 \sqrt{|\S|}   \right) &  \left(  2c_2   +1-   \frac{\epsilon}{\gamma} c_1
  -   \beta \right) \nonumber \\
&\hspace{1cm}<
\sqrt{d} \left( 1 -   \frac{\epsilon}{\gamma} c_1
  -   \xi \right)
\label{eq:condonabsvalthm}
\end{align}
where $c_1,c_2,c_3$, and $c_4$ are constants depending on $\delta_{\max}(\S)$ and $\mu_{\max}(\S)$ only. Then, for $\xi>0$ such that $\max\{1-16 e, \alpha^2 \} < \xi^2 \leq \alpha^2 (1+16e)$, with probability at least 
\begin{align}
1  -  \exp \! \left(   -  \d   \frac{  (\xi^2-\alpha^2)^2}{512 e^2  \gamma^2  \alpha^2 } \right) 
\label{eq:probestnoismn}
\end{align} 
the solution to $\POPT$ applied to $\y^{(i)} = \A^{(i)} \x^{(i)}, i=0,...,\d-1$,  and denoted by $\tilde \x^{(0)},...,\tilde \x^{(\d-1)}$, is supported on $\S$ and satisfies  
\begin{align} 
\left( \sum_{i=0}^{\d-1} \norm[2]{\tilde \x^{(i)} -\x^{(i)}}^2 \right)^{1/2}
\leq c_3 \epsilon  + \gamma \,  c_4  \sqrt{|\S|}.
\label{eq:defxstSinthm}
\end{align}
\label{thm:mnnoisy}
\end{theorem}
The main implication of Theorem \ref{thm:mnnoisy} is that, under certain conditions on the $\A^{(i)}$ and for the noise level $\epsilon$ sufficiently small, the probability that $\POPT$ produces a solution with correct support set that is ``close'' in $\ell_2$-norm to the true $\x^{(i)}$, tends  to $1$ exponentially fast in $\d$. This result is also new for the MMV case.  
Condition \eqref{eq:condonabsvalthm} ensures that the noise level $\epsilon$ is sufficiently small. Note that \eqref{eq:condonabsvalthm} depends on the ``worst'' measurement matrix through $\delta_{\max}(\S)$ and  $\mu_{\max}(\S)$. This is sensible as noise has the largest effect on the measurement $\y^{(i)}$ taken through the ``worst'' measurement matrix. 

\addtolength{\textheight}{-6cm} 

We finally turn to the performance of MOMP. This result will be stated for i.i.d.~sub-Gaussian $\x_\S^{(i)}$.

\begin{theorem}
Fix $\S \subseteq \{0,...,n-1\}$ with cardinality $\sparsity\defeq |\S|$, and let the measurement matrices $\A^{(0)}, ..., \A^{(\d-1)}\in \reals^{m\times n}$  have unit norm columns with $\mu_{\max}(\S)<1$ and $\delta_{\max}(\S) <1$. Let the entries of $\x^{(0)}_\S\!,...,\x^{(\d-1)}_\S  \in \reals^{\sparsity}$ be  i.i.d.~zero-mean $\rho$-sub-Gaussian with unit variance. 
Suppose that 
\begin{align}
\epsilon \leq \frac{1-\delta_{\max}(\S) }{ 1-\delta_{\max}(\S) + (1-\delta_{\max}(\S) ) \mu_{\max}(\S) } \varkappa
\label{eq:ompbon}
\end{align}
for some $\varkappa\geq 0$. 
If
\begin{align}
&\sqrt{1-\beta}  \left(  \frac{1}{d} \sum_{i=0}^{\d-1}  \left( 1-   \frac{\mu_i^2(\S)}{1-\delta_i(\S)}  \right)^2 \right)^{1/2} \nonumber \\ 
&\hspace{1cm}  -  \sqrt{1+\beta} \left( \frac{1}{d} \sum_{i=0}^{\d-1}  \left(\frac{\mu_i(\S)}{1-\delta_i(\S)} \right)^2 \right)^{1/2}  \geq \varkappa
\label{eq:ompavreccond}
\end{align}
for $\beta$ satisfying $0<\beta \leq 32 e \rho$, then with probability at least \eqref{eq:bocorrrecprob}, MOMP applied to $\y^{(i)} = \A^{(i)} \x^{(i)}, i=0,...,\d-1$, yields an estimate of the $\x^{(i)}$, denoted by $\tilde \x^{(i)}$, that is supported on $\S$ and satisfies  
\begin{align} 
\left( \sum_{i=0}^{\d-1} \norm[2]{\tilde \x^{(i)} -\x^{(i)}}^2 \right)^{1/2}
 \leq \frac{1+\delta_{\max}(\S)}{1-\delta_{\max}(\S)} \epsilon.
\label{eq:ompnoisbodiff}
\end{align} 
\label{thm:ompavcase_noisy}
\end{theorem}

Again, the main implication of Theorem \ref{thm:ompavcase_noisy} is that, under certain mild conditions on the $\A^{(i)}$ and for the noise level $\epsilon$ sufficiently small, the probability that MOMP produces a solution with correct support set that is ``close'' to the true $\x^{(i)}$, tends to $1$ exponentially fast in $\d$. This was shown in \cite{gribonval_atoms_2008} for the MMV case and for i.i.d.~Gaussian $\x_\S^{(i)}$. Note that for $\epsilon =0$, i.e., in the noiseless case, \eqref{eq:ompavreccond} reduces to \eqref{eq:ompavreccond_nc} and Theorem \ref{thm:ompavcase_noisy} reduces to Theorem \ref{thm:ompavcase}. For $\epsilon > 0$, and hence $\varkappa>0$, 
\eqref{eq:ompavreccond} is more restrictive than Condition \eqref{eq:ompavreccond_nc}. Condition \eqref{eq:ompbon} depends on the ``worst'' measurement matrix, and ensures that the noise level $\epsilon$ is sufficiently small. 
The constants in Theorem \eqref{thm:ompavcase_noisy} can be improved for i.i.d.~Gaussian $\x_\S^{(i)}$ \cite{heckel_generalized_2012}. 

We conclude by noting that the results in this paper extend straightforwardly to the case of complex $\A^{(i)}$ and $\x^{(i)}$.

\vspace{0.3cm}


\end{document}